\documentclass[aps,prl,twocolumn,showpacs,superscriptaddress,twoside,showpacs]{revtex4-2}
\usepackage{graphicx}
\usepackage{dcolumn}
\usepackage{bm}
\usepackage{booktabs}
\usepackage{amsmath}
\usepackage{color}

\begin{document}

\title{Comment on ``First Observation of Multiple Transverse Wobbling Bands of Different Kinds in $^{183}$Au [Phys. Rev. Lett. 125, 132501 (2020)]"}

 \author{S. Guo}
\affiliation{CAS Key Laboratory of High Precision Nuclear Spectroscopy, Institute of Modern Physics, Chinese Academy of Sciences, Lanzhou 730000, China} 
\affiliation{School of Nuclear Science and Technology, University of Chinese Academy of Science, Beijing 100049, People's Republic of China}
 \author{C. M. Petrache}
\affiliation{Universit\'{e} Paris-Saclay, CNRS/IN2P3, IJCLab, 91405  Orsay, France}

 \date{}
\begin{abstract}

In [S. Nandi et al., Phys. Rev. Lett. 125, 132501 (2020)] two transverse wobbling bands were reported in $^{183}$Au. The critical experimental proof for this assignment is the $E2$ dominated linking transitions between the wobbling and normal bands, which are supported by fitting the measured DCO ratio and polarization results. However, the uncertainties are significantly underestimated according to an analysis on the statistical error. With reasonable error, the mixing ratios cannot be exclusively decided, and the $M1$ dominated character cannot be excluded, indicating that the wobbling interpretation is questionable.

\end{abstract}

\pacs{21.10.Re, 21.60.Ev, 23.20.Lv, 27.60.+j}

\keywords{ Nuclear reaction: linear polarization measurement}

\maketitle

This comment quests on the reliability of the reported transverse wobbling bands in Ref. \cite{Nandi} via an analysis of the statistical error.

To deduce the mixing ratio \cite{PDCO}, the polarization ($P$) is extracted using the formula

\begin{equation}
	P = A/Q = \frac{a C_\perp-C_\parallel}{a C_\perp+C_\parallel}/Q,\label{e1}
\end{equation}

\noindent where $C_\perp$ and $C_\parallel$ denote the number of coincidence counts between the segments of the clover detector in the direction perpendicular and parallel to the emission plane, respectively. Polarization sensitivity ($Q$) and asymmetry correction ($a$) were two necessary calibrations absent in Ref. \cite{Nandi}.

\begin{figure}[ht]
	\vskip -. cm
	\hskip -. cm
	\centering\includegraphics[clip=true,width=0.3\textwidth]{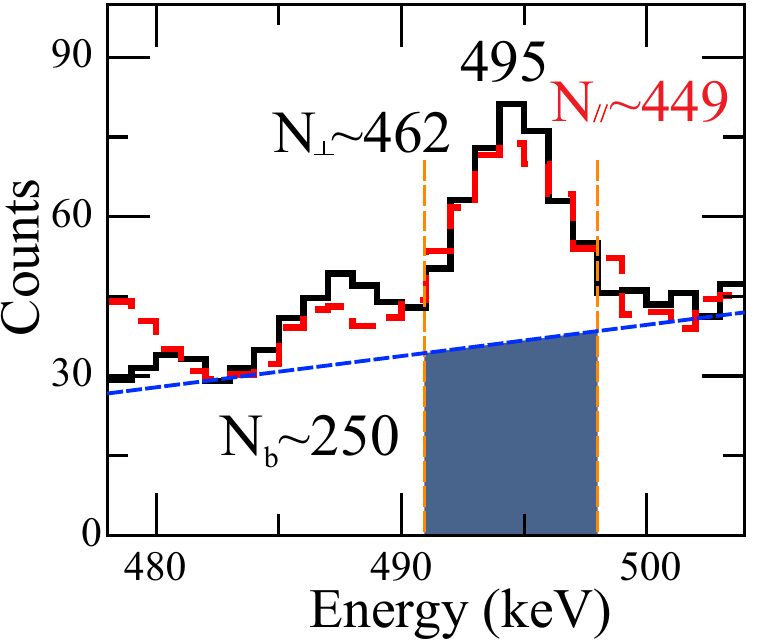}
	\vskip -0. cm
	\caption{(Color online) Partial $\gamma$-ray spectra gated by 502-keV transition extracted from Fig. 1c of Ref. \cite{Nandi}. The counts are obtained by summing the areas between the orange lines. The background is estimated as shown in blue. }
	\label{fig1}
\end{figure}

Here we examine the statistical error on the $P$ of the 495-keV transition, based on the corresponding spectra shown in Fig. \ref{fig1}. Surprisingly apparent backgrounds exist below the 495-keV peaks in both perpendicular and parallel spectra.


According to the spectra,

\begin{equation}
	\begin{aligned}
		C_\perp &= N_\perp - N_b \sim 212,\\
		C_\parallel &= N_\parallel - N_b \sim 199.\label{e2}
	\end{aligned}
\end{equation}

One should not neglect the errors induced by subtracting the spectra gated by the Compton platform near to the gating transition (marked by an extra subscript ``$b$"), from that gated by the peak channels of the gating transition (marked by an extra subscript ``$p$"). Therefore the counts can be further written as

\begin{equation}
	\begin{aligned}
		N_\perp &= N_{\perp p} - N_{\perp b} \sim 462,\\
		N_\parallel &= N_{\parallel p} - N_{\parallel b} \sim 449,\\
		N_b &= N_{b p} - N_{b b} \sim 250.\label{e3}
	\end{aligned}
\end{equation}

Their statistical errors are

\begin{equation}
	\begin{aligned}
		\sigma(N_\perp) &= \sqrt{462+2 \sigma^2(N_{\perp b})},\\
		\sigma(N_\parallel) &= \sqrt{449+2 \sigma^2(N_{\parallel b})},\\
		\sigma(N_b) &= \sqrt{250+2 \sigma^2(N_{b b})}.\label{e4}
	\end{aligned}
\end{equation}

According to error propagation formulas in textbooks such as Ref. \cite{Knoll}, 

\begin{equation}
	\sigma_u \approx \sqrt{\frac{\sigma_x^2+\sigma_y^2}{(x+y)^2}}, (u=\frac{x-y}{x+y}, x \approx y).\label{e5}
\end{equation}

Assuming $a=1$ and imposing $Q_{495}$ = 0.233 (using the calibration in Ref. \cite{PDCO}) into Eqs. \ref{e1}-\ref{e5},

\begin{equation}
	\begin{aligned}
		\sigma_P(495) &\approx \frac{\sqrt{\sigma^2 (C_{\perp})+\sigma^2 (C_{\parallel})}}{(C_{\perp}+C_{\parallel})Q}\\
		&= \frac{\sqrt{1411 + 2\sigma^2(N_{\perp b}) + 2\sigma^2(N_{\parallel b}) + 4\sigma^2(N_{b b})}}{(212+199)0.233}.\label{e6}
	\end{aligned}
\end{equation}

As a conservative estimation, one can assume that the errors on $N_{\perp b}$, $N_{\parallel b}$, and $N_{b b}$ are all around $\sqrt{N_b}$ ($\sqrt{250}$), leading to $\sigma_P(495) \approx 0.61$, which is much larger than the underestimated reported error $\sigma_P \approx 0.09$. According to the PDCO curves plotted in Fig. 2 of Ref. \cite{Nandi}, it is impossible to exclude one of the two solutions with an error of $\approx 0.61$, and therefore to definitely establish the character of the 495-keV transition.

In addition, the wobbling nature to the $h_{9/2}$ band was also supported by the deduced mixing ratio of another 498-keV connecting transition. However, the contamination from another adjacent 498-keV $E2$ transition cannot be excluded in the spectrum gated by any known transition.

In summary, the uncertainties of the polarization results were significantly underestimated in the commented work, leading to the lack of solid experimental supports for the wobbling interpretation.

This work has been partly supported by the National Natural Science Foundation of China, under contract No. U1932137.

\end{document}